\documentstyle[11pt,newpasp,twoside,graphicx]{article}
\index{pulsars}
\index{bubbles}
\index{supernova remnants}

\def\edcomment#1{\iffalse\marginpar{\raggedright\sl#1\/}\else\relax\fi}
\reversemarginpar

\begin{document}
\title{Some arguments in 
support of the association of PSR B\,1706$-$44 with the supernova remnant 
G\,343.1$-$2.3}

\author{Douglas C.-J.\ Bock}
\affil{Radio Astronomy Laboratory, University of California at 
Berkeley, 601 Campbell Hall, Berkeley, CA 94720, USA; 
dbock@astro.berkeley.edu}
\author{V.\ V.\ Gvaramadze}
\affil{E.K.Kharadze Abastumani Astrophysical Observatory, Georgian Academy of
Sciences, A. Kazbegi ave. 2-a, Tbilisi, 380060, Georgia;
vgvaram@mx.iki.rssi.ru}

\begin{abstract}
  We present some arguments in support of the association of the
  pulsar PSR B\,1706$-$44 with the supernova remnant G\,343.1$-$2.3, based
  on the idea that these objects could be the result of a supernova
  explosion within a mushroom-like cavity (created by the supernova
  progenitor wind breaking out of the parent molecular cloud). We
  suggest that in addition to the known bright ``half" of 
  G\,343.1$-$2.3 there should exist a more extended and weaker
  component, such that the actual shape of G\,343.1$-$2.3 is similar
  to that of the well-known SNR VRO\,42.05.01. We have found such a
  component in archival radio data.

\end{abstract}

\section{Introduction}
%
PSR B\,1706$-$44 (Johnston et al.\ 1992) is superposed on the outer
edge of an incomplete arc of radio emission discovered by McAdam,
Osborne, \& Parkinson (1993). McAdam et al.\ interpreted the arc as a
shell-type supernova remnant (SNR), named G\,343.1$-$2.3, and
suggested that it is physically associated with PSR B\,1706$-$44. This
suggestion was questioned by Frail, Goss, \& Whiteoak (1994) and
Nicastro, Johnston, \& Koribalski (1996) on three bases: Gaensler
\& Johnston's (1995) statistical study, which suggests that young
($<25\,000$ yr) pulsars cannot overrun their parent SNR shells (the
spin-down age of PSR B\,1706$-$44 is $\simeq 17\,500$ yr); an
inconsistency between the implied and measured (scintillation)
transverse velocities of the pulsar; and the absence of any
apparent interaction between the pulsar and the SNR's ``shell".  In
this paper we show how the existing observational data on PSR
B\,1706$-$44 and G\,343.1$-$2.3 can be interpreted in favor of their
physical association (cf.\ Dodson et al.\ 2001).

\pagebreak

\section{The supernova remnant G\,343.1$-$2.3}

\subsection{Observational data}

The 843 MHz image of G\,343.1$-$2.3 by McAdam et al.\ (1993) shows a 
well-defined arc (a half-ellipse) of radio emission of maximum extent 
about $40^{'}$. A VLA image of the SNR obtained by Frail 
et al.\ (1994) shows the existence of weak, diffuse emission both 
inside and outside the bright arc. This emission fills a region similar 
to and about two times more extended than the bright arc (Dodson et al.\ 
2001; see also Duncan et al.\ 1995 and Fig.\ 2).

\subsection{The origin of the pulsar/SNR system}

\begin{figure}
\centering
\includegraphics[width=\textwidth]{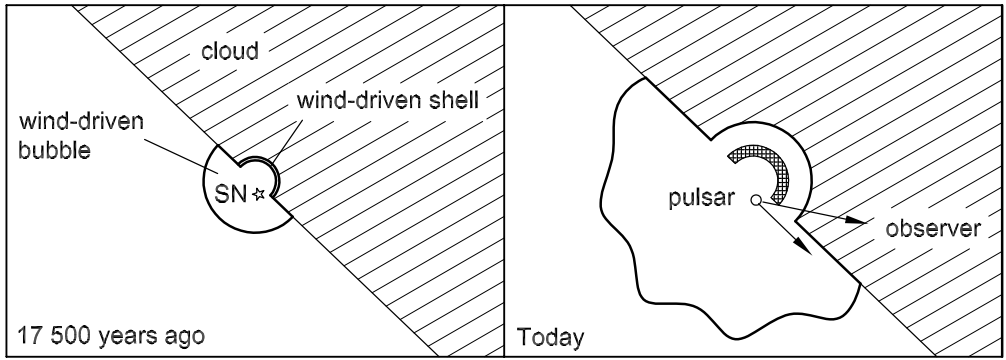}
\caption{Schematic of the proposed origin of G\,343.1$-$2.3}
\end{figure}

We suggest that the SNR G\,343.1$-$2.3 is the result of an
off-centered cavity supernova (SN) explosion. Fig.\ 1 schematically
depicts a scenario for its origin. A massive star (the progenitor of
the SN) ends its evolution within a mushroom-like cavity formed by the
SN progenitor wind breaking out of the parent molecular cloud and
expanding into an intercloud medium of much less density. The proper
motion of the progenitor star results in a considerable offset of the
SN explosion site from the geometrical center of the semi-spherical
cavity created inside the cloud; we suggest that the SN exploded
outside the cloud. The subsequent interaction of the SN blast wave
with the reprocessed ambient medium determines the structure of the
resulting SNR (e.g.\ Franco et al.\ 1991), which acquires a form
reminiscent of the well-known SNR VRO\,42.05.01.  We speculate that
the wind-blown cavity formed inside the cloud was surrounded by a
shell of mass less than some critical value (for spherically-symmetric
shells this value is about 50 times the mass of the SN ejecta; e.g.\ 
Franco et al.\ 1991), so that the SN blast wave was able to overrun
the shell to propagate further into the unperturbed gas of the cloud,
leaving behind the reaccelerated and gradually broadening turbulent
shell.
We suggest that the bright arc of
G\,343.1$-$2.3 corresponds to the shocked former wind-driven shell and that
the diffuse radio emission comes from the ``half" of the SN blast wave
propagating into the molecular cloud (see Fig.\ 1).
These two components of the SNR form the ``stem" of the ``mushroom".
We expect that a more extended component of the SNR (the ``cap" of the
``mushroom") should exist to the southeast of the bright arc, 
which corresponds
to the ``half" of the SN blast wave expanding in the intercloud medium.
It is curious that the 2.4 GHz Parkes Survey (Duncan et al.\ 1995)
reveals such a structure (Fig.\ 2). Although this structure
could be a foreground or background object, its location in the ``proper" 
place and its symmetry with
respect to the bright arc of G\,343.1$-$2.3 suggests that it could
be physically related to this SNR. 

\begin{figure}
\centering
\includegraphics[width=9cm]{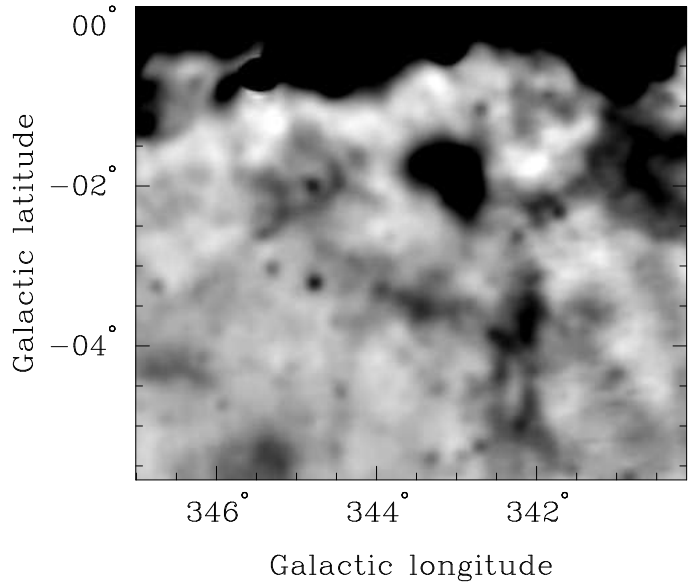}
\caption{2.4~GHz image of G\,343.1$-$2.3 (Duncan et al.\ 1995)}
\end{figure}

\section{Reliability of the pulsar/SNR association}

\subsection{On the statistical argument against the association}

Although it is now clear that PSR B\,1706$-$44 is located (at least in
projection) well within the SNR G\,343.1$-$2.3 (Dodson et al.\ 2001),
we note here that Gaensler \& Johnston (1995) did not consider
two very important effects: modification of
the ambient medium by the ionizing emission and stellar wind of massive
stars (the progenitors of most SNe), and the proper motion of SN progenitor
stars (Gvaramadze 2000, 2002). Taking into account these two effects allows it to be shown
that even a young pulsar moving with a moderate 
velocity ($\simeq 200\, {\rm km}\,{\rm s}^{-1}$) is able to escape the
SNR's shell, provided it was born not far from the edge of the
wind-driven bubble. Alternatively, the apparent location
of a pulsar on the edge of SNR's shell could be due simply to the effect
of projection in non-spherically-symmetric SNRs (see Fig.\ 1).

\subsection{On the pulsar velocity}

The implied pulsar transverse velocity, i.e.\ the velocity inferred
from the angular displacement of PSR B\,1706$-$44 from the geometrical
center of the (bright) arc, is $ V_{\rm imp} \simeq 700 \, \theta
_{20} D_{2.1} \tau _{17.5} ^{-1} \, {\rm km}\,{\rm s}^{-1} ,$ where
$\theta _{20}$ is the angular displacement in units of $20^{''}$,
$D_{2.1}$ is the distance to the pulsar in units of 2.1 kpc, and $\tau
_{17.5}$ is the spin-down age of the pulsar in units of 17.5 kyr.
Nicastro et al.\ (1996) compared this estimate to one derived from
scintillation measurements, finding the latter anomalously low
($\simeq 0.05 V_{\rm imp}$).
The inconsistency was used by Nicastro et al.\ to suggest that the
pulsar did not originate from the apparent center of SNR, and that the
pulsar and SNR are not associated. We agree with their first
suggestion (see \S2.2) and therefore believe that the implied velocity
can be reduced.  On the other hand we have found (Bock \& Gvaramadze,
in preparation) that if the turbulent material of the reaccelerated
former wind-driven shell (the bright arc of SNR) is responsible for
nearly all the scattering of PSR B\,1706$-$44, then the pulsar moves in the
same direction and with nearly the same (transverse) velocity as does
the part of the bright arc projected on the pulsar (cf.\ Gvaramadze
2001).  If so, one can use the non-detection of soft X-ray
emission from the SNR (Becker, Brazier, \& Tr\"umper 1995) to show
that the pulsar transverse velocity should indeed be less than $V_{\rm
  imp}$, though it can be much larger than that calculated by Nicastro
et al.\ (1996).  We also expect that the pulsar proper motion is from
the northeast to the southwest; this should be tested observationally.

\subsection{On the absence of interaction between the pulsar
and the SNR}

The absence of any morphological signature of an interaction between PSR
B\,1706$-$44 and G\,343.1$-$2.3, despite the apparent proximity of the
pulsar to the bright arc of the SNR, can easily be explained if the SN
indeed exploded within a mushroom-like wind-driven cavity, as
discussed above.

\acknowledgments

We are grateful to R.\ Dodson for providing his manuscript in advance of 
publication. VVG is grateful to the LOC for financial support.

\end{document}